\documentstyle[aaspptwo]{article}
\tighten
\begin{document}
\title{On the Number Density of Sunyaev--Zel'dovich 
Clusters of Galaxies.}

\author{C. Hern\'andez--Monteagudo, 
F. Atrio--Barandela}

\affil{F{\'\i }sica Te\'orica. Universidad de Salamanca. 37008 Spain.\\
	email: chm@orion.usal.es, atrio@astro.usal.es}

\author{J. P. M\"ucket}

\affil{Astrophysikalisches Institut Potsdam. D-14482 Potsdam.\\
 	email: jpmuecket@aip.de}

\begin{abstract}
If the mean properties of 
clusters of galaxies are well described by the
entropy-driven model, the distortion induced by the cluster
population on the blackbody spectrum of the Cosmic
Microwave Background radiation is proportional to
the total amount of intracluster gas 
while temperature anisotropies are
dominated by the contribution of $10^{14}$M$_\odot$
clusters. This result depends marginally on
cluster parameters and it can be used to estimate the number
density of clusters with 
enough hot gas to produce a detectable Sunyaev-Zel'dovich
effect. Comparing different cosmological models, the 
relation depends mainly on the density parameter $\Omega_m$.
If the number density of clusters could be estimated by a
different method, then this dependence could 
be used to constrain $\Omega_m$. 

\end{abstract}


\section {Introduction.}
Clusters of galaxies can be detected in the X-ray band 
due to the emission of the intracluster (IC) gas.
This  gas is hot enough to change the brightness
of the Cosmic Microwave Background (CMB) photons through
inverse Compton scattering (Zel'dovich \& Sunyaev 1969).
The Sunyaev-Zel'dovich (SZ)
effect caused by individual clusters has been measured 
for tens of clusters (see Birkinshaw 1999 for a review).
Eventually, the PLANCK satellite will produce an all-sky
catalogue likely to contain thousands of SZ sources
(da Silva et al. 1999). Together with the effect of
single sources, the overall cluster population
induces distortions and temperature anisotropies on the CMB.
The Sunyaev-Zel'dovich effect has a 
well known frequency dependence that helps
to distinguish it from other foregrounds
and methods have already been
proposed to separate and measure this contribution 
(Hobson et al. 1998).

Extensive theoretical work has been devoted to
analyze the effect of clusters on the
CMB radiation (Cole \& Kaiser 1988, 
Bartlett \& Silk 1994, Colafrancesco et al. 1994, 
Atrio-Barandela \& M\"ucket 1999 -hereafter paper I-,
Komatsu \& Kitayama, 1999). Barbosa et al. (1996) 
noticed that the value of the mean
Comptonization parameter, that measures the amplitude
of the blackbody distortion, depends on which are the
less massive clusters that produce
a significant effect. On the other hand,
Cole \& Kaiser (1988) remarked
that temperature anisotropies were dominated by massive
clusters at moderate redshifts. Since clusters of a given 
mass contribute differently to temperature anisotropies
and distortions,  in this letter we show how
the number density of clusters with enough hot gas to
produce a detectable SZ effect can be estimated.
This number is marginally dependent
on the cluster model and it varies by at most
a factor of four in different cosmologies.

\section{Scaling relations of clusters.\label{scaling}}

Following paper I, we assume that clusters are spherical
in shape, with typical size the virial radius $r_v$, 
virial mass $M$ and have virialized at redshift $z$.
We also assume that the IC gas is isothermal,
distributed smoothly and with a
radial distribution well fitted
by the spherical isothermal $\beta$ model
$n_e(r) = n_c[1+(r/r_c)^2]^{-3\beta/2}$
(Jones \& Forman 1984),
where $n_c, r_c$ are the central electron density and
the core radius of the cluster, respectively.
For convenience, we shall assume that $\beta = 2/3$.
Similar results to those presented here are obtained
for any value of $\beta$ within the observed range
($0.5\le\beta\le 0.7$, see Markevitch et al. 1997).
Further, we assume that the
dynamical evolution of the IC gas is
well described by the entropy-driven model by Bower (1997)
since this  model seems to provide an adequate description of clusters
with redshifts $z\le 0.4$ (Mushotzky \& Scharf 1997). 

To compute the spectral distortion and 
temperature anisotropies induced by hot
IC gas we need to translate the properties of  a sample of clusters
at low redshifts into their equivalents at earlier
epochs. For the spherical collapse
model, the virial radius scales with mass
and redshift as $r_v= r_{vo} (M/10^{15}M_\odot)^{1/3}(1+z)^{-1}$.
If the electron temperature is proportional to the velocity
dispersion of the dark matter then
$T  = T_g (M/10^{15}M_\odot)^{2/3}(1+z)$.
In the entropy-driven model, the electron central
density scales as
$n_c = n_{co} (T/T_g)^{3/2} (1 + z)^{-3\epsilon/2}$.
Finally, the core radius scales as
$r_c = r_{co} (M/10^{15}M_\odot)^{-1/6} (1 + z)^{(-1+3\epsilon)/4}$.
In this scaling relations, $r_{co}, r_{vo}, T_g, n_{co}$ are the current
average core and virial radius, the central 
gas temperature and electron density of a $10^{15}$M$_\odot$ cluster,
respectively, while $\epsilon$ parametrizes the rate 
of core entropy evolution.  
Mushotzky and Scharf (1997) found $\epsilon = 0\pm 0.9$.

\section{Distortions and Temperature anisotropies.}

The effect of the IC gas
is both to distort the blackbody spectrum
and to induce temperature anisotropies on the CMB.
A measure of such distortion is the 
mean Comptonization parameter defined as
\begin{equation}
\bar y = g(hf/K_BT_o) \int {dn\over dM}dM{dV\over dz}dz \kappa y_o \bar\phi .
\label{ycmean}
\end{equation}
In this expression $g(x) = x{\rm coth}(x/2) - 4$
gives the dependence of  the SZ effect with frequency
$f$; $T_o$ is the CMB mean temperature,
$dn/dM$ is the cluster number density
per unit of mass, $y_o = [k_B\sigma_T/m_ec^2]r_c T_g n_c$
with physical constants having their usual meaning and
$\kappa=(r_v/2d_A)^2$ gives the probability that a particular
line of sight crosses a cluster, with $d_A$ the angular
distance to the cluster. If we introduce
$p=r_v/r_c$ then $\bar\phi=4p^{-2}(p-\tan^{-1}p)$ 
is the averaged line of sight through a cluster.

The effect of all clusters on an angular scale $l$ can
be obtained by adding in quadrature the contribution
of each single cluster. If we assume that clusters
are Poisson distributed on the sky the radiation
power spectrum is (Cole \& Kaiser 1988)
\begin{equation}
P(l)    = \int {dn\over dM}dM{dV\over dz}dz
        (g(x)y_o)^2|\tilde \phi(l)|^2,
        \label{pq}
\end{equation}
where $\tilde \phi(l)$ is the Fourier transform of the angular
profile of the IC gas.  This expression represents the
contribution of the foreground cluster population to
the power spectrum of CMB temperature anisotropies.
The variance of the temperature field is given by
\begin{equation}
<({\Delta T\over T_o})^2>= {1\over 2\pi}\int ldl W(l) P(l)
\label{dtt}
\end{equation}
In this equation, $W(l)$ represents the window function of the
experiment.  This temperature anisotropy, together with other
secondary contributions,
adds in quadrature with the "intrinsic" (cosmological)
signal to give the pattern of temperature anisotropies
on the sky (see paper I for details).

To compute the number density of clusters as a function 
of mass and redshift we
use the Press-Schechter approximation, known to describe
accurately the abundance of clusters in numerical simulations
at different redshifts
\begin{equation}
{dn\over dM} =\sqrt{2\over\pi}\big({\rho\over M}\big)
{\nu\over\sigma} {d\ln\sigma \over dM} e^{-1/2(\nu/\sigma)^2} .
\label{ps}
\end{equation}
In this expression,
$\nu = (\delta_vb/\sigma_8)$ is the peak-height
threshold, $\sigma$ is the rms mass fluctuation within
a top-hat filter measured in units of $\sigma_8$, 
the rms mass fluctuation on a 
sphere of $8h^{-1}$Mpc today, $\delta_v$ is the threshold 
overdensity of the spherical
collapse model and $b$ is the bias factor. 

\section{The cluster number density.}

We integrate numerically eqs. (\ref{ycmean}) and (\ref{dtt})
for three cosmological models with the same
baryon fraction $\Omega_B = 0.05$: two flat models, (1)
standard CDM, with 
$\Omega_{cdm} = 0.95$,  (2) $\Lambda$CDM with 
$\Omega_{cdm} = 0.25$ and $\Omega_\Lambda = 0.7$
as indicated by the recent measurements of high redshift
supernovae (Perlmutter et al. 1999) and (3) an open model,
OCDM, with $\Omega_{cdm} = 0.25$ and no cosmological
constant,  consistent with cluster dynamics and
evolution (Bahcall 1999). As average parameters, for
a $10^{15}$M$_\odot$ cluster we took: $r_{vo} = 1.3 h^{-1}$Mpc,
$T_g = 10^8$K, $r_{co} = 0.13h^{-1}$Mpc and 
$n_{co} = 1.19\cdot 10^{-3} h^{1/2}$cm$^{-3}$.
We considered different gas evolutionary histories,
by taking $\epsilon = -1, 0, 1$.
The matter power spectrum was normalized to
$\sigma(8h^{-1}$Mpc$)=0.7$ (Einasto et al. 1999b)
and we considered the following bias factors: $b =1.05, 1.3$.
The first corresponds to the bias of QDOT galaxies and
the second is an average value obtained from several catalogues 
(Einasto et al 1999a). Also, we considered $\delta_v = 1.5, 1.7$
(Tozzi \& Governato, 1998,  Governato et al. 1999).
The simplicity of the Press-Schechter 
approach lies on the fact that the number
density depends only on the ratio $\nu = \delta_vb/\sigma_8$.
With the values quoted above, $\nu$ varies in the range
2 to 3 and we shall quote our results for both limits.

To compute the effect of the cluster population on
the CMB radiation it is necessary to determine first if
groups of galaxies of about $10^{13}$M$_\odot$
have enough hot gas to produce a significant contribution.
The scaling relations of Sec.~[2]
have been found to be accurate for clusters
above $10^{14}$M$_\odot$. However, they could
very well be extended below this mass scale.
In this respect, it is necessary to specify 
the integration range of eqs.~(\ref{ycmean}) and (\ref{pq}).
We tried three different lower mass limits:
$M_{min} = 0.1, 0.5$ and $1$ (in 
units of $10^{14}$M$_\odot$). The main conclusion
was that when changing the mass limit one order of magnitude
$\bar y$ could vary a factor 3 to 10, 
depending on the model,
but the temperature anisotropy $\Delta T/T_o$
varied by less than 30\%. In this calculations  and
in the rest of the paper, temperature anisotropies were computed using 
the window function of the SuZie experiment (Church et al. 1997).
For the models considered above, the anisotropies
in the Rayleigh-Jeans regime
ranged from $14\mu$K for sCDM with $\nu=2$ to
$1\mu$K for $\Lambda$CDM with $\nu=3$.

Integration of eq~(\ref{pq}) shows that $l^2P(l)$
reaches a maximum  around $l=1000-3000$.
At those scales Komatsu \& Kitayama (1999) have
concluded that the increase in temperature anisotropy
induced by cluster correlations is negligible.
To illustrate the different behavior of
temperature anisotropies and distortions with
the lower mass integration limit we shall consider the
behavior of the radiation power spectrum at 
$l=1000$. This scale is close to the maximum of the SZ effect and
is a good compromise between 
the smallest angular scale that will be resolved by PLANCK
and the maximum of the SuZie window function.
In Fig. 1 we show $d\bar y/dM, dP(l=1000)/dM$ for different 
cosmological models. 
Notice that the integrand of $\bar y$  increases
at the low mass end, but that of 
the radiation power spectrum reaches a maximum
around $10^{14}$M$_\odot$. Its exact location is
weakly dependent on model parameters.
A qualitative explanation of Fig. 1 can be given:
since $\sigma\sim 1$ on a mass scale of 
$\approx 5\times 10^{14}$M$_\odot$, in the upper
mass limit $d\bar y/dM$ and $dP/dM$ are damped
by the exponential factor in the Press-Schechter
formula; in the low mass limit,
the exponential factor is close to unity and
$d\bar y/dM \sim M^{(1+n)/6}$ but
$dP/dM \sim M^{n/6+3/2}$ where $n$ is the slope
of the matter power spectrum at  $10^{13}$M$_\odot$,
roughly $n\simeq -2$ (Einasto et al. 1999a).

To conclude, the gas in 
groups contributes significantly to 
Comptonization but little
to temperature anisotropies. This fact can be used to estimate
the number density of clusters that  produce a significant
SZ effect. For this purpose, in paper I we introduced
the parameter $\eta = {\Delta T/T_o \over \bar y}$.
It represents a relation similar to 
the approximate solution of the full kinetic equation for the
change of photon distribution due to inverse Compton scattering
for a single cluster: $\Delta T/T_o = g(hf/kT_o)y_o$.
As an example, in Fig. 2a we plot $\eta$ with respect to
lower mass integration limit for three cosmological
models with $\epsilon = 0$: solid line (sCDM), dot-dashed line (OCDM)
and dashed line ($\Lambda$CDM). 
Thick lines correspond to $\nu = 3$ and thin lines $\nu = 2$.
Similar behavior can be observed for other gas
evolution histories and model parameters.

By inspection of eqs.~(\ref{ycmean}) and (\ref{pq}),
the Comptonization parameter and the power spectrum
scale linearly with the number 
density of clusters. Therefore,
one expects that $\eta$ will scale as $n_{cl}^{-1/2}$.
In fig. 2b we plot $\eta n_{cl}^{1/2}$ for the same models
as in Fig. 2a.  Notice that the value of $\eta n_{cl}^{1/2}$
is almost constant and nearly independent on the threshold $\nu$. 
The actual value varies by a factor of two when the 
total matter content changes from
$\Omega_{m} = 1$ to 0.3 but does not depend on the
geometry of the cosmological model, i.e. on the value 
of $\Lambda$. In Fig. 3 we plot $\eta$ as a function of
gas evolution history, assuming that the number density 
of all the objects contributing to the SZ effect is $2.5\times 
10^{-4}h^3$Mpc$^{-3}$. In each panel, thick 
lines correspond to a threshold $\nu = 3$ and
thin lines to $\nu = 2$. Solid lines correspond to
a lower mass integration limit 
of $10^{14}$M$_\odot$ and dot-dashed lines to
$10^{13}$M$_\odot$. Notice that $\eta$ varies by
less than $30\%$ for different gas evolutionary histories.
Let as remark that, if $\Omega_m = \Omega_B + \Omega_{cdm}$ is known,
$\eta$ permits to obtain $n_{cl}$. Vice versa, if 
$n_{cl}$ could be determined by other means, 
$\Omega_m$ could be determined.  If no assumption is made
about the cosmology, $n_{cl}$ can be determined within
a factor of four.

\section{Discussion.}

The results of previous section indicates that
$\eta$ is mostly sensitive to the number
density of clusters, and for a given cosmological
model is weakly dependent on the exact modeling
of the cluster population. Therefore,
a measurement of both the
mean distortion of the CMB radiation and the temperature
anisotropies induced by clusters of galaxies will permit
to estimate the number density of objects that
have enough hot gas to produce a detectable effect
on the CMB. 
This conclusion is not limited by the validity of
the  scaling relations of Sec.~[2].
One must expect a similar behavior for a different cluster
model, even though the level of
temperature anisotropies and distortions will be different. 
If the hot IC gas was uniformly distributed in the Universe, it
would still produce blackbody distortions but will not give
rise to temperature anisotropies, and $\eta$ would
be zero. Only when the gas is clumped
in clusters, two different lines of sight, seeing different
column densities, will have different temperatures.
Therefore, $\bar y$ scales with the number of clusters
along the line of sight while for Poisson
distributed clusters $\Delta T/ T_o$ scales as the square root 
and $\eta\propto n_{cl}^{-1/2}$, as shown.
Therefore, independently of the exact
model used to describe the average cluster properties,
$\eta$ measures how the hot gas is distributed
in the Universe.  Finally, if the cluster number density
is known, $\eta$ could be used as an estimator of
the total matter density $\Omega_m$.

To summarize, clusters of $10^{14}$M$_\odot$
mass dominate the contribution to temperature anisotropies.
Therefore, even if the adequacy of the scaling relations of
Sec.~[2]  below that mass scale has not been established,
temperature anisotropies can be reliably calculated
using eqs.~(\ref{pq}) and (\ref{dtt}). If
the entropy-driven model proves not to be adequate to describe 
groups of galaxies, one would have to use 
hydrodynamical simulations such as those of da Silva et al.
(1999) to compute $\bar y$ but 
$\eta$ would still provide an estimate of the
number density of SZ clusters. Comparison of this
density with that of X-ray clusters will undoubtably
help our understanding of cluster formation and evolution.

{\bf Acknowledgments.}
We thank the referee, J. Einasto, for very constructive comments
and suggestions.
This research was supported by Spanish German Integrated Actions
HA 97/39. FAB would like to acknowledge the support of the
Junta de Castilla y Le\'on, grant SA40/97.

\clearpage

\begin{figure*}[t]
\plotone{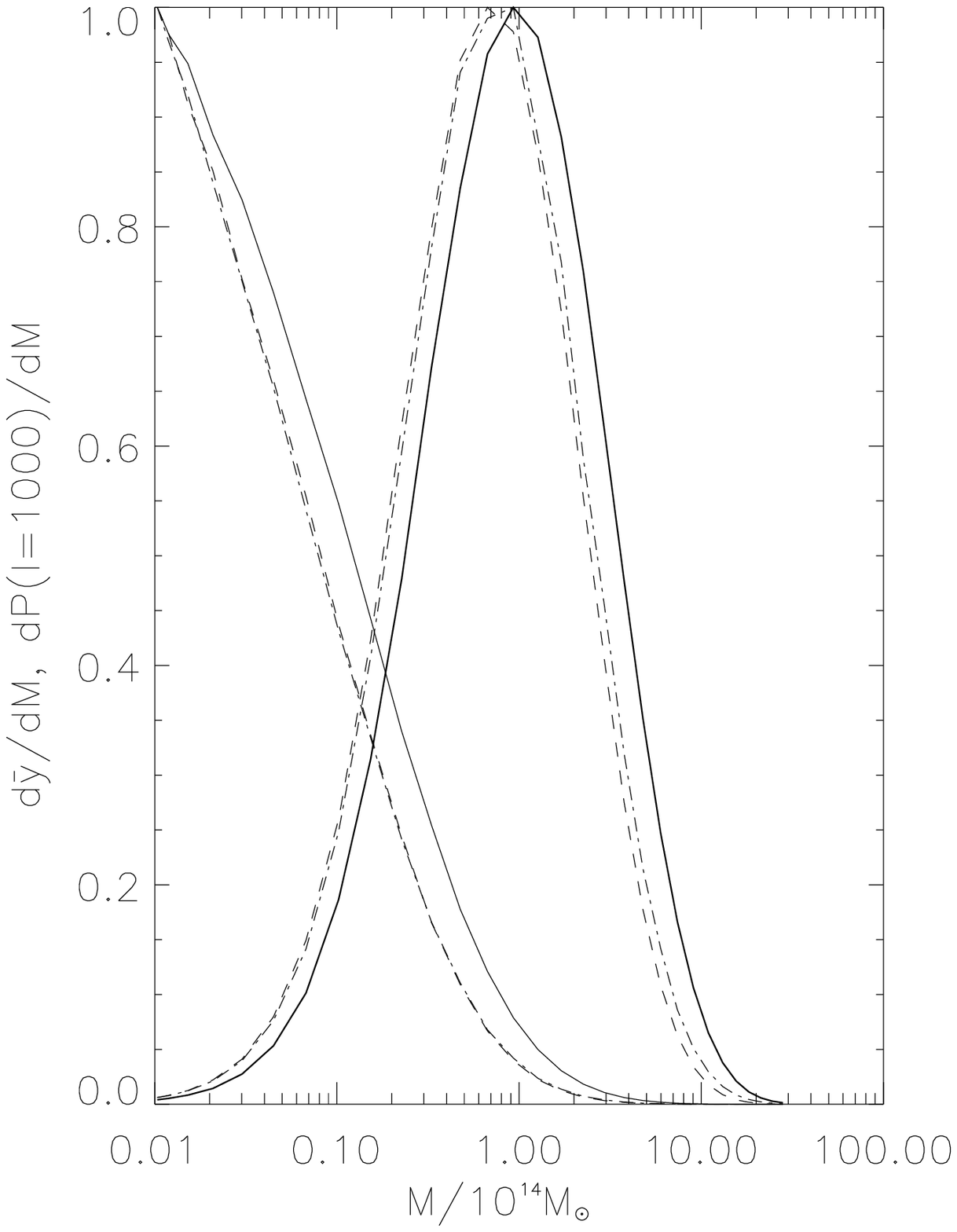}
\caption[FIG. 1]{
Integrand of the mean Comptonization parameter (curves on the left
side) and the radiation power spectrum (on the right)
for different models. Solid lines correspond to
standard CDM, dashed lines to $\Lambda$CDM and dot-dashed line to OCDM.
The scale in the y-axis is arbitrary; the data has been
rescaled for convenience. The amplitudes for $d\bar y/dM$
at $10^{12}$M$_\odot$ and $dP/dM$ at the maximum are:
$3\times 10^{-4}$, $1.1\times 10^{-16}$ for sCDM, 
$1.5\times 10^{-4}$, $1.4\times 10^{-17}$ for OCDM and
$1.3\times 10^{-4}$, $1\times 10^{-17}$ for $\Lambda$CDM.
All these quantities are in units of $(M/10^{15}$M$_\odot)^{-1}$
and we took $g(x) = 1$.
}
\label{fig1}
\end{figure*}

\clearpage

\begin{figure*}[t]
\plotone{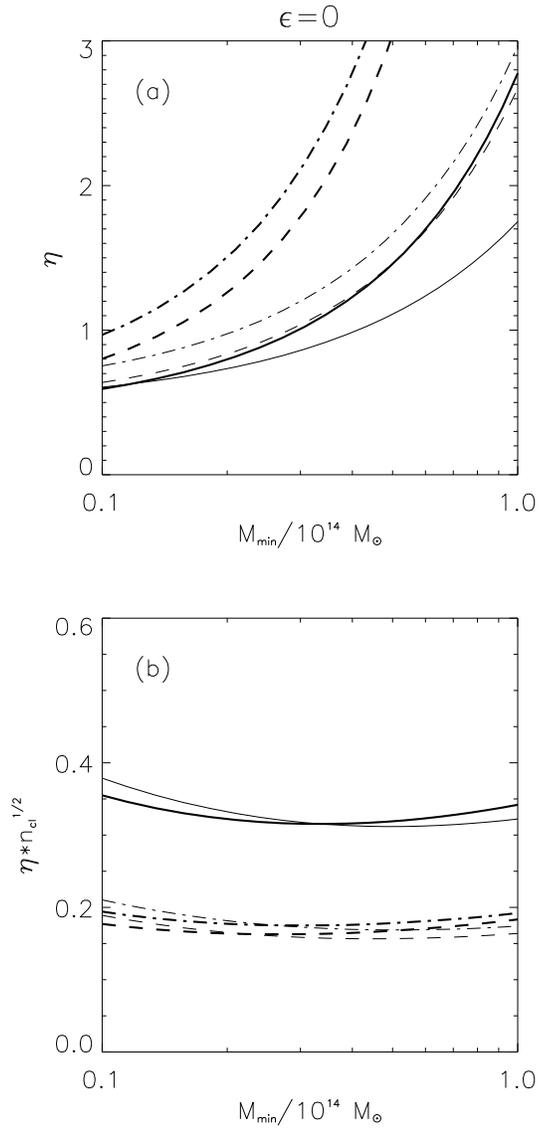}
\caption[FIG. 2]{
(a) Variation of $\eta$ (three upper curves) with the lower mass
limit of integration $M_{min}$ for three different cosmological
models: SCDM (solid line), OCDM (dot-dashed line) and
$\Lambda$CDM (dashed line). Thick lines correspond to 
a threshold $\nu = 3$ and thin lines to $\nu = 2$.
(b) Variation of $\eta n_{cl}^{1/2}$ a function of $M_{min}$
for the same cosmological models as in (a).
}
\label{fig2}
\end{figure*}

\clearpage

\begin{figure*}[t]
\plotfiddle{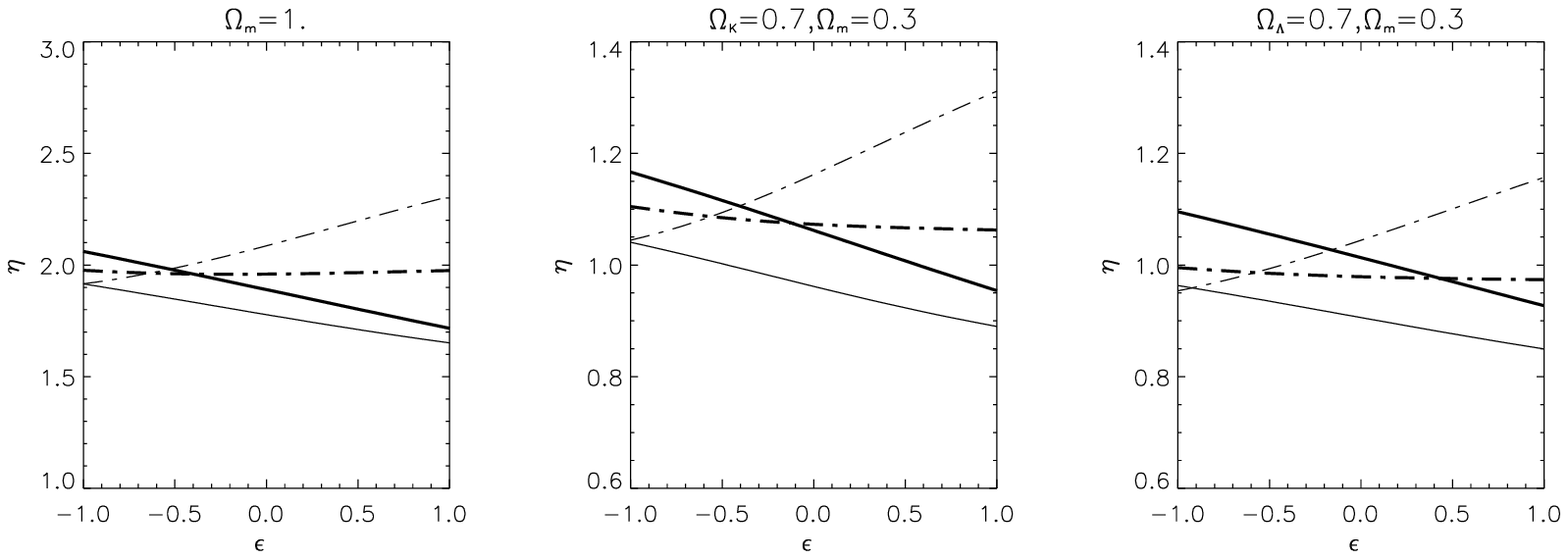}{2cm}{0.0}{90.0}{90.0}{-200.0}{+10.0}
\caption[FIG. 3]{
The value of $\eta$ rescaled to a cluster density 
of $2.5\times 10^{-4} h^3$Mpc$^{-3}$ clusters, as a function
of the gas evolutionary history. From left
to right: sCDM, OCDM and $\Lambda$CDM. 
Thick lines correspond to $\nu = 3$ and thin lines
to $\nu = 2$; solid lines to $M_{min} = 1$
and dot-dashed lines to  $M_{min} = 0.1$
(in units of $10^{14}$M$\odot$).
}
\label{fig3}
\end{figure*}


\begin{references}

\reference{} Atrio--Barandela, F. \&  M\"ucket, J. 1999, \apj, 515, 465
	(paper I)
\reference{} Bahcall, N. A. 1999 in "Proceedings of the American
	Astronomical Society Meeting", 194, 3002
\reference{} Barbosa, D., Bartlett, J. G., Blanchard, A., Oukbir, J.
	1996, A\& A, 314, 13
\reference{} Bartlett, J.G., Silk, J. 1994, \apj, 423, 12
\reference{} Birkinshaw, M. 1999, Phys. Rep., 310, 97
\reference{} Bower, R.G. 1997, MNRAS, 288, 355
\reference{} Colafrancesco, S., Mazzotta, P., Rephaeli, Y. \&
	Vittorio, N. 1994, \apj, 433, 454
\reference{} Cole, S. \& Kaiser, N. 1989, MNRAS, 233, 637
\reference{} Church, S. E. et al. 1997 \apj, 484, 523
\reference{} da Silva, A. C., Barbosa, D., Liddle, A. R. 
	\& Thomas, P. A. 1999, preprint astro-ph/9907224
\reference{} Einasto, J. et al. 1999a, \apj, 519, 441
\reference{} Einasto, J. et al. 1999b, \apj, 519, 456
\reference{} Governato et al. 1999, MNRAS, submitted. Preprint
	astro-ph/9810189
\reference{} Hobson, M. P., Jones, A. W., Lasenby, A. N., 
	Bouchet, F. R. 1998, \mnras, 300, 1
\reference{} Jones, C. \& Forman, W. 1984, \apj, 276, 38
\reference{} Komatsu, E. \& Kitayama, T. 1999, preprint astro-ph/9908087
\reference{} Markevitch, M., Forman, W.R., Sarazin, C.L. \& Vikhlinin, A. 
	1997, \apj, 503, 77
\reference{} Mushotzky, R.F. \& Scharf, C.A. 1997, \apj, 482, L13
\reference{} Perlmutter, S. et al. 1999, \apj, 517, 565
\reference{} Press, W.H., \& Schechter, P. 1974, \apj, 187, 425
\reference{} Tozzi, P. \& Governato, F. 1998, in "The Young Universe"
S. D'Odorico, A.  Fontana, and E. Giallongo, eds. ASP Conference Series, 
146, 461
\reference{} Zel'dovich, Ya. B., Sunyaev, R. 1969, \apss, 4, 301 
\end{references}
\end{document}